\documentclass[
    ,final            
  ,numberedheadings 
  ]
  {aipproc}

\layoutstyle{6x9}

\def\be{\begin{equation}}
\def\ee{\end{equation}}
\def\bea{\begin{eqnarray}}
\def\eea{\end{eqnarray}}
\def\beq{\begin{equation}}
\def\eeq{\end{equation}}

\def\mk{{\mathbf k}}%

\begin{document}

\title{$\phi$$\phi$ Back-to-Back Correlations in Finite Expanding
Systems}


\author{Sandra S. Padula}{
  address={Inst. de F\'{\i}sica Te\'orica,
UNESP~-~Rua Pamplona 145, 01405-900 S\~ao Paulo, SP - Brazil}, }

\author{Y. Hama}{address={Instituto de F\'{\i}sica, USP, Caixa Postal 66318, 05389-970 S\~{a}o Paulo, SP -
Brazil}, }

\iftrue
\author{G. Krein}{address={Inst. de F\'{\i}sica Te\'orica,
UNESP~-~Rua Pamplona 145, 01405-900 S\~ao Paulo, SP - Brazil},  }

\author{P. K. Panda}{address={Depto. de F\'\i sica-CFM, UFSC - C. P. 476, 88040-900 Florian\'opolis, SC -
Brazil}, }

\author{T. Cs\"org\H o}{address={MTA KFKI RMKI, H - 1525 Budapest 114, P.O. Box 49, Hungary}}

\begin{abstract}
Back-to-Back Correlations (BBC) of particle-antiparticle pairs are
predicted to appear if hot and dense hadronic matter is formed in
high energy nucleus-nucleus collisions. The BBC are related to
in-medium mass-modification and squeezing of the quanta involved.
Although the suppression of finite emission times were already
known, the effects of finite system sizes and of collective
phenomena had not been studied yet. Thus, for testing the survival
and magnitude of the effect in more realistic situations, we study
the BBC when mass-modification occurs in a finite sized,
thermalized medium, considering a non-relativistically expanding
fireball with finite emission time, and evaluating the width of
the back-to-back correlation function. We show that the BBC signal
indeed survives the expansion and flow effects, with sufficient
magnitude to be observed at RHIC.
\end{abstract}

\maketitle

%
%

In the late 1990's the {\sl Back-to-Back Correlations} ({\sl BBC})
between boson-antiboson pairs were shown to exist if the particles
masses were modified in a hot and dense medium\cite{acg99}. Not
much longer after that, it was also shown that a similar BBC
existed between fermion-antifermion pairs with medium-modified
masses\cite{pchkp01}. A similar formalism is applicable to both
BBC cases, related to the Bogoliubov-Valatin transformations of
in-medium and asymptotic operators. Both the bosonic (bBBC) and
the fermionic (fBBC) Back-to-Back Correlations are positive and
have unlimited magnitude, thus differing from the
identical-particle correlations, also known as HBT (Hanbury Brown
\& Twiss) correlations, which are limited for both cases, being
negative in the fermionic sector. BBC were expected to be
significant for $p_T < 2$ GeV/c. Nevertheless, already in the
Ref.\cite{acg99}, it has been shown that the duration of the
emission process significantly suppresses its magnitude. The
effects of finite system sizes and of collective phenomena were
not known, which motivated us to investigate their consequences.
Some preliminary results will be discussed here and illustrated in
some particular cases.


\vspace{-0.5cm}
\subsection{Infinite Homogeneous Medium}

In our analysis, we assume the validity of local thermalization
and  hydrodynamics, from the beginning up to the system
freeze-out, as well as a short duration of the particle emission.
We also consider that the effective in-medium Hamiltonian can be
written as
\be {H}={H}_0-\int d{\bf x} d{\bf y} \phi({\bf x}) \delta M^2({\bf
x}-{\bf y})\phi({\bf y})
 , \label{H}
\label{e:ham} \ee
where
\be {H}_0 =  \int d{\bf x}({\dot{\phi}}^2 + |{\bf \nabla}\phi|^2+
m^2 \phi^2)  \label{H0} \ee
is the asymptotic (free) Hamiltonian in the matter rest frame, and
the second term in Eq. (\ref{e:ham}) describes the medium
modifications. The scalar field $\phi$ represents quasi-particles
propagating with momentum-dependent medium-modified mass
$m_\star$, related to the vacuum mass, $m$, by
$$
m_\star^2(|{\bf k}|)=m^2-\delta M^2(|{\bf k}|).
$$
This implies that the dispersion relations in the vacuum and
in-medium are given, respectively, by $\omega_k^2 = m^2+{\bf k}^2
\nonumber$ and $\Omega_k^2 = m_\star^2+{\bf k}^2=\omega_k^2 -
\delta M^2(|{\bf k}|)$,
where $\Omega$ is the frequency of the in-medium mode with
momentum ${\bf k}$.

The annihilation (creation) operator, $b$ ($b^\dagger$), for the
in-medium, thermalized quasi-particles is related to the
annihilation (creation) operator of the asymptotic, observed
quanta with momentum $k^\mu = (\omega_k,{\bf k})$, $a$
($a^\dagger$), by the Bogoliubov-Valatin transformation
\be a^\dagger_k=c^*_k b^\dagger_k + s_{-k} b_{-k}  \; ; \; a_k=c_k
b_k + s^*_{-k} b^\dagger_{-k}, \label{bogliu}\ee
where $c_k=\cosh(f_k)$ and $s_k=\sinh(f_k)$; $f_k=\frac{1}{2}
\log(\frac{\omega_k}{\Omega_k})$ is the so-called {\sl squeezing
parameter}, since the Bogoliubov transformation is equivalent to a
squeezing operation.

On the other hand, the two-particle probability distribution is
given by \be \langle a^\dagger_{\mk_1} a^\dagger_{\mk_2} a_{\mk_2}
a_{\mk_1} \rangle = \Bigl[\langle a^\dagger_{\mk_1}
a_{\mk_1}\rangle \langle a^\dagger_{\mk_2} a_{\mk_2} \rangle +
\langle a^\dagger_{\mk_1} a_{\mk_2}\rangle \langle
a^\dagger_{\mk_2} a_{\mk_1} \rangle + \langle a^\dagger_{\mk_1}
a^\dagger_{\mk_2} \rangle \langle a_{\mk_2} a_{\mk_1}
\rangle\Bigr].
 \label{rand} \ee

The first term on the r.h.s. is proportional to the product of the
two single-inclusive distributions, with momenta $k_i$, i.e.,  $
N_1(\mk_i) = \omega_{\mk_i} \frac{d^3N}{d\mk_i} = \omega_{\mk_i}\,
\langle a^\dagger_{\mk_i} a_{\mk_i} \rangle $, while the second
term is proportional to the square modulus of $
G_c({\mk_1},{\mk_2}) = G_c(1,2) = \sqrt{\omega_{\mk_1}
\omega_{\mk_2}} \; \langle a^\dagger_{\mk_1} a_{\mk_2} \rangle$,
the so-called {\bf chaotic amplitude}. The last term is related to
this new contribution, $ G_s({\mk_1},{\mk_2}) = G_s(1,2) =
\sqrt{\omega_{\mk_1} \omega_{\mk_2}} \; \langle a_{\mk_1}
a_{\mk_2} \rangle $, which is
called 
{\bf squeezed amplitude}.

In cases where the particle is its own anti-particle (for
$\pi^0$$\pi^0$ or $\phi$$\phi$ boson pairs, for instance), both
terms contribute, and the full correlation function is written as
\be C_2({\mk_1},{\mk_2}) = 1 +
\frac{|G_c({\mk_1},{\mk_2})|^2}{G_c({\mk_1},{\mk_1})
G_c({\mk_2},{\mk_2})} + \frac{|G_s({\mk_1},{\mk_2})
|^2}{G_c({\mk_1},{\mk_1}) G_c({\mk_2},{\mk_2}) }, \label{fullcorr}
\ee where the first two terms correspond to the HBT correlation,
and last term, represents this additional contribution to the
correlation function, i.e., the squeezing part.

In the above equations, the amplitudes were written in terms of
the asymptotic operators. However, the averages in Eq.
(\ref{rand}) are estimated by means of the density operator
$\hat{\rho}$, as the thermal average for globally thermalized gas
of $b$ quanta that is homogeneous in the system with a certain
volume V, with $\hat{\rho}=(1/Z)\exp[-(V/T)(1/(2\pi)^3) \int
\Omega b^\dagger b]$. Being so, the above equations should be
expressed in terms of the in-medium operators by means of Eq.
(\ref{bogliu}), prior to performing the thermal averages.

We know that the maximum value of the HBT correlation function is
attained when $k_1=k_2=k$, resulting in $C_c(k,k)=2$. Accordingly,
the maximum value of the BBC correlation function is attained for
$k_1=-k_2=k$, resulting, after performing the thermal averages,
into
\be C_s(\mk, -\mk)  =  1 + \frac{|c^*_{\mk} s_{\mk} n_{\mk} +
c^*_{-\mk} s_{-\mk} \left(n_{-\mk} + 1\right)|^{\, 2} } {n_1(\mk)
\, n_1(-\mk)}, \label{e:c2b} \ee
where $n_1(\mk) = \left[|c_{\mk}|^2 n_{\mk} + |s_{-\mk}|^2
\left(n_{-\mk} + 1 \right)\right]$ is related to the spectral
function by $ N_1(\mk) = \frac{V}{(2 \pi)^3} \, \omega_\mk \,
n_1(\mk)$; $n_\mk$ is the Bose-Einstein distribution function of
the in-medium quanta with energy $\Omega_\mk$ at temperature $T$.
Strictly speaking Eq.~(\ref{e:c2b}) is valid only in the rest
frame of the medium.

\vspace{-0.5cm}
\subsection{Finite size medium moving with collective velocity}

For a hydrodynamical ensemble, both the chaotic and the squeezed
amplitudes, $G_c$ and $G_s$, respectively, can be written in the
special form derived by Makhlin and Sinyukov~\cite{MakSyniukov}
(see Eqs.~(22) and (23) of Ref.~\cite{acg99}), namely
\be G_c({\mk_1},{\mk_2}) = \int \frac{d^4\sigma_{\mu}(x)}{(2
\pi)^3} \, K^\mu_{1,2} \, e^{i \, q_{1,2} \cdot x} \,
\Bigl\{|c_{1,2}|^2 \, n_{1,2}(x) + |s_{-1,-2}|^2 \,
\left[n_{-1,-2}(x) + 1\right] \Bigr\}, \label{e:gc} \ee \be
G_s({\mk_1},{\mk_2})  = \int \frac{d^4\sigma_{\mu}(x)}{(2 \pi)^3}
\, K^\mu_{1,2} \, e^{2 \,i \, K_{1,2} \cdot x} \Bigl\{
s^*_{-1,2}\, c_{2,-1} \, n_{-1,2}(x) + c_{1,-2} \, s^*_{-2,1} \,
\left[n_{1,-2}(x) + 1\right] \Bigr\}. \nonumber\\
\label{e:gs} \ee

In Eq. (\ref{e:gc}) and (\ref{e:gs}),
$d\sigma^4_\mu(x)=d^3\Sigma_\mu(x,\tau)F(\tau)d\tau$ is the
product of the normal-oriented volume element depending
parametrically on $\tau$ (freeze-out hyper-surface parameter) and
on its invariant distribution, $F(\tau)$; $\sigma^\mu$ is the
hydrodynamical freeze-out surface.

For studying the expansion of the system we adopt the
non-relativistic hydrodynamical model of Ref.~\cite{Csorgo:fg}. In
this model the fireball expands in a spherically symmetric manner
with a local flow vector given by the four-velocity $u^\mu =
\gamma \, (1, {\mathbf v})$, assumed to be non-relativistic, with
$\gamma = (1-{\mathbf v}^2)^{-1/2} \approx 1 +{\mathbf v}^2/2$,
where
$${\mathbf v} =
\langle u\rangle {\mathbf r}/R,$$ being $\langle u\rangle$ and $R$
the mean expansion velocity and the radius of the fireball,
respectively. We then divide the inhomogeneous medium into
independent cells and assume that the expressions for $G_c$ and
$G_s$ can be evaluated locally within each cell using
Eq.~(\ref{bogliu}). The squeezing coefficient can be written, in
more detail, as
\begin{equation}
f_{i,j}(x)=\frac{1}{2}\log\left[\frac{K^{\mu}_{i,j}(x)\, u_\mu
(x)} {K^{*\nu}_{i,j}(x) \, u_\nu(x)}\right] = \frac{1}{2}
\log\left[\frac{\omega_{\mk_i}(x) + \omega_{\mk_j}(x)}
{\Omega_{\mk_i}(x) + \Omega_{\mk_j}(x)}\right] \equiv f_{\pm i,\pm
j}(x), \label{e:rxk}
\end{equation}
where, as in HBT, the pair momentum difference and the pair
average momentum are given, respectively, by $%
q^{\mu}_{i,j}(x) = k^{\mu}_i(x) - k^{\mu}_j(x)$, and
$K^{\mu}_{i,j}(x) = \frac{1}{2} \left[k^{\mu}_i(x) + k^{\mu}_j(x)
\right]$.
In addition, we consider the Boltzmann limit of the Bose-Einstein
distribution for $n_k$, i.e., $n_{i,j}\sim \exp{[-(K^\mu_{i,j}
u_\mu - \mu(x))/T(x)]}$, and assume a time-dependent parametric
solution of the hydrodynamical equations, i.e.,
$\mu(x)/T(x)=\mu_0/T_0 - r^2/(2R^2)$, as in Ref. \cite
{{Csorgo:fg}}. Furthermore, we consider two possible scenarios for
the freeze-out: the first, corresponds to a sudden freeze-out, in
which $F(\tau)\propto \delta(\tau-\tau_0)$. The second scenario
corresponds to a smeared freeze-out, for which
$\frac{\theta(\tau-\tau_0)}{\Delta \tau}e^{-(\tau-\tau_0)/\Delta
\tau}$. This last more realistic scenario has a dramatic effect on
the Back-to-Back Correlation function, as already showed in
Ref.\cite{acg99}, by reducing severely the signal's magnitude,
even for a smearing of about $\Delta \tau \sim 2$ fm/c.

In discussing finite-size effects, we distinguish between the
volume of the entire thermalized medium, denoted by $V$ (with
radius $R$), and the volume filled with mass-shifted quanta,
denoted by $V_s$ (with radius $R_s$). Naturally, $V_s \leq V$ in
the general case. In the derivation of the expressions for
$G_c(1,2)$ and $G_s(1,2)$, for simplicity, we introduce a Gaussian
profile function in the integrands, i.e., $\sim \exp{[- {\bf
r}/(2R)^2]}$. In Fig. 1 we illustrate this by showing
cross-sectional areas corresponding to Gaussian profiles, for the
cases with $V=V_s$ and $V>V_s$.

\begin{figure}[t]
  \includegraphics[height=.2\textheight]{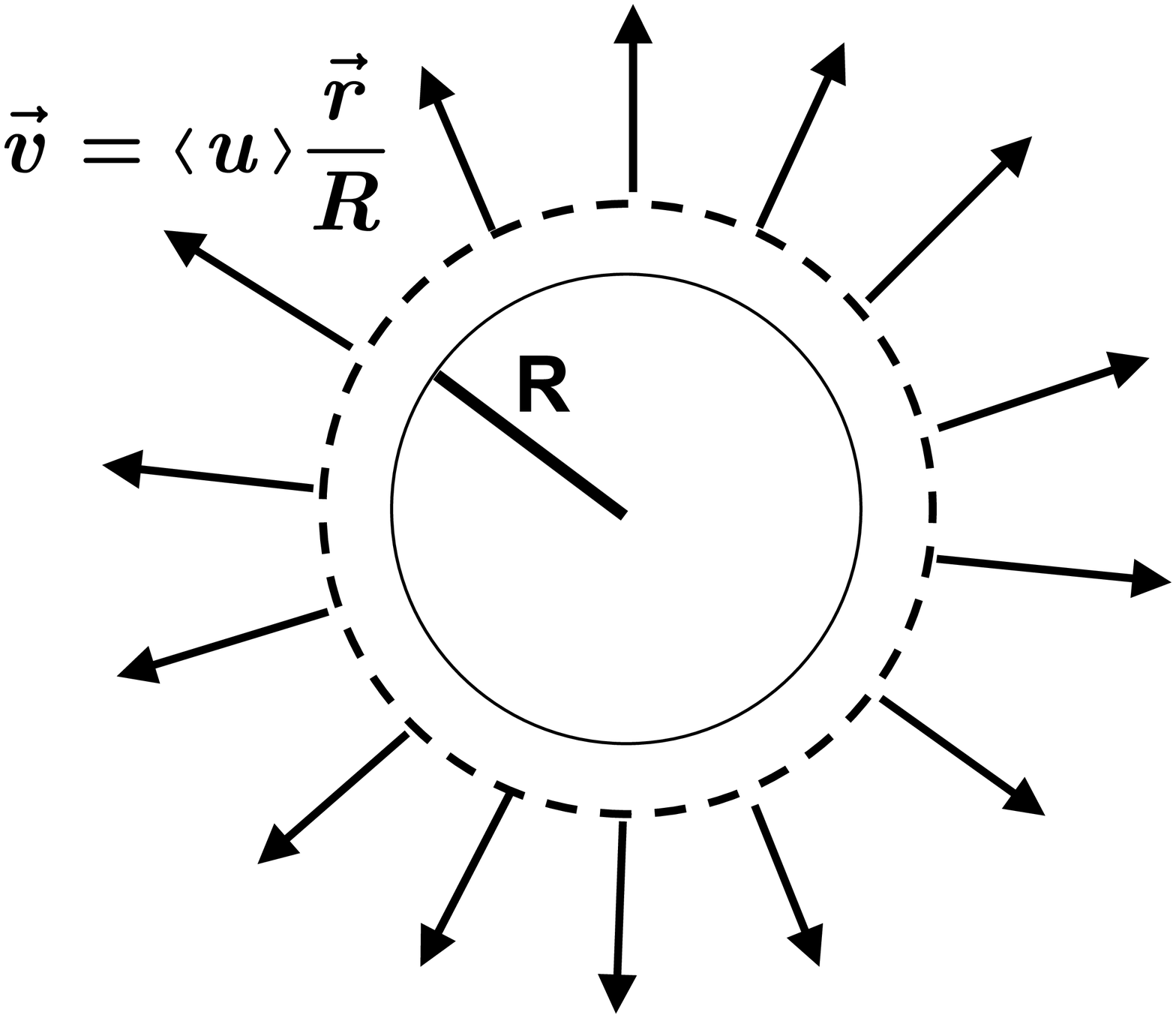}
  \includegraphics[height=.2\textheight]{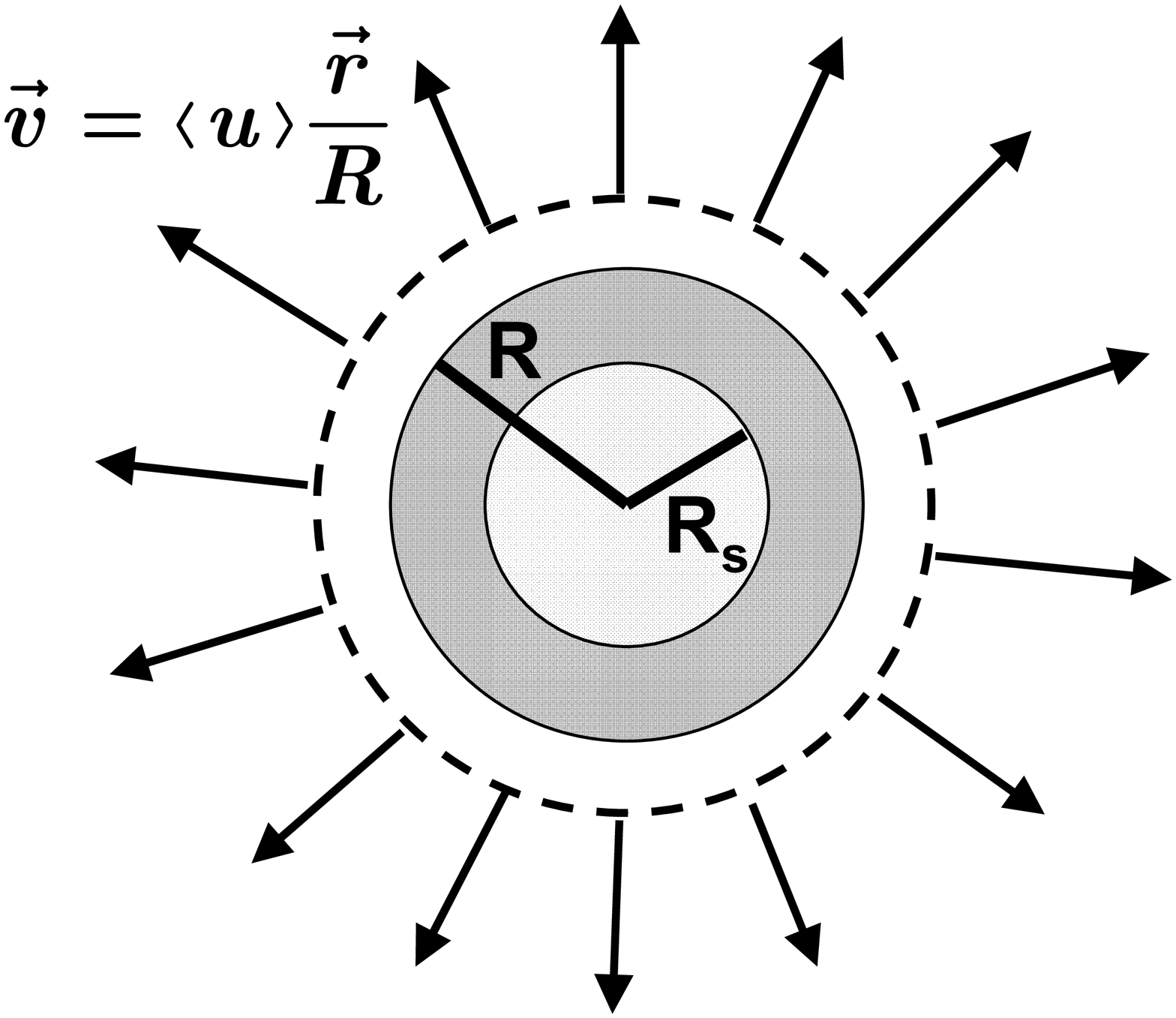}
 \caption{Schematic representation of the region where the mass-shift
 occurs: on the l.h.s., the modification is extended to the whole system, whereas
 on the r.h.s. it happens only in a smaller portion of the thermalized medium. The figures
 represent cross-sectional areas of the Gaussian profiles.}
\end{figure}

In the non-relativistic limit, the accounting for the squeezing
effects can be simplified for small mass shifts $(m_\star - m)/m
\ll 1$, such that the squeezing parameter in Eq. (\ref{e:rxk}) can
be written simply as $ f(i,j,{\mathbf r}) \approx
\frac{1}{2}\log\left(\frac{m}{m_\star}\right)$. This limit is
important, because in this case the coordinate dependence enters
the squeezing parameter $f$ only through the possible position
dependence of the mass-shift which, in principle, could be
calculated from thermal field models in the local density
approximation. Therefore, in an approximation such that the
position dependence of the in-medium mass can be neglected, the
$c(i,j)=c_0$ and $s(i,j)=s_0$ factors can be removed from the
integrands in Eq. (\ref{e:gc}) and (\ref{e:gs}) and all what
remains to be done are Fourier transforms of Gaussian functions.


\begin{figure}[t]
  \includegraphics[height=.33\textheight]{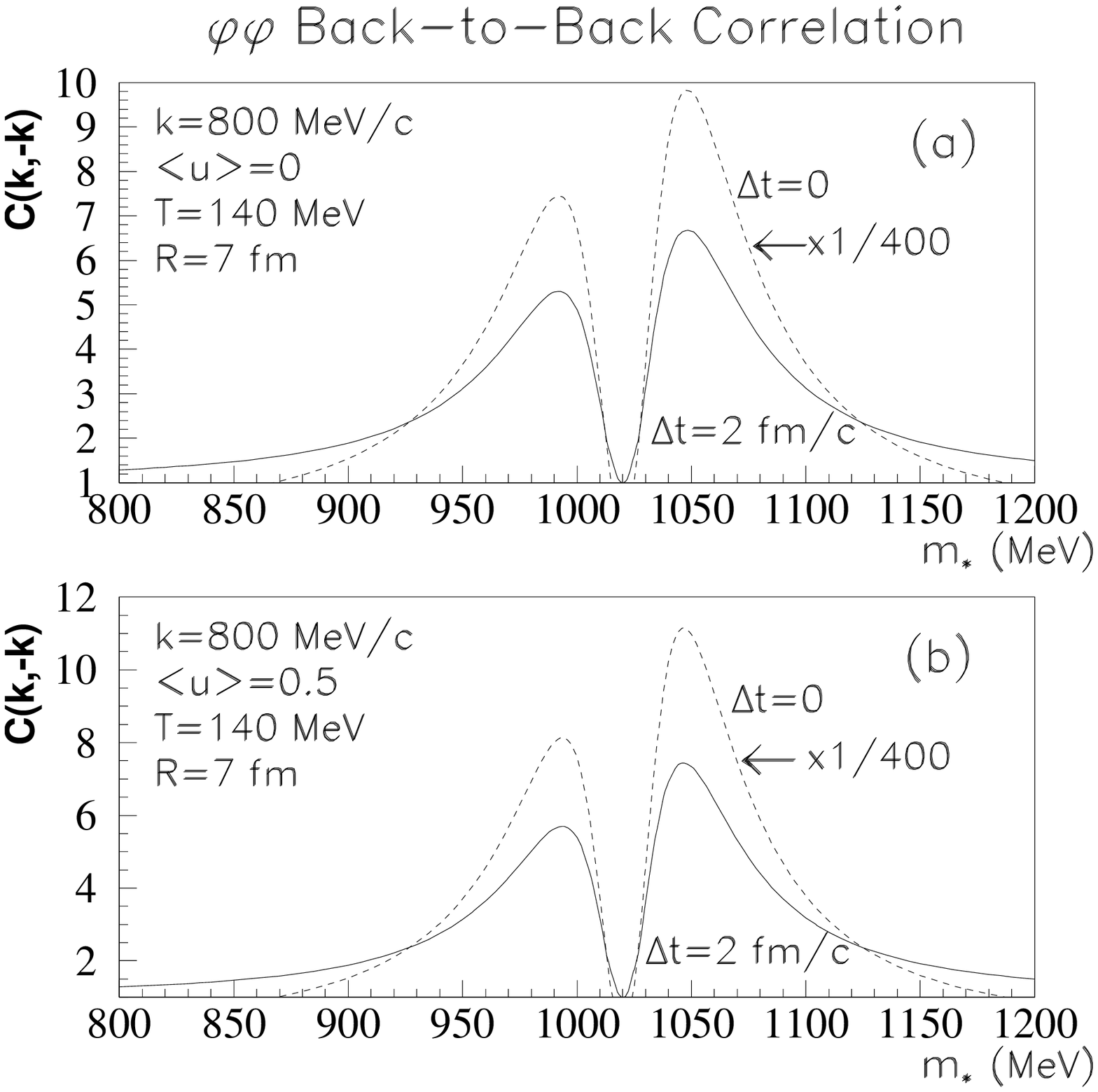}
  \includegraphics[height=.33\textheight]{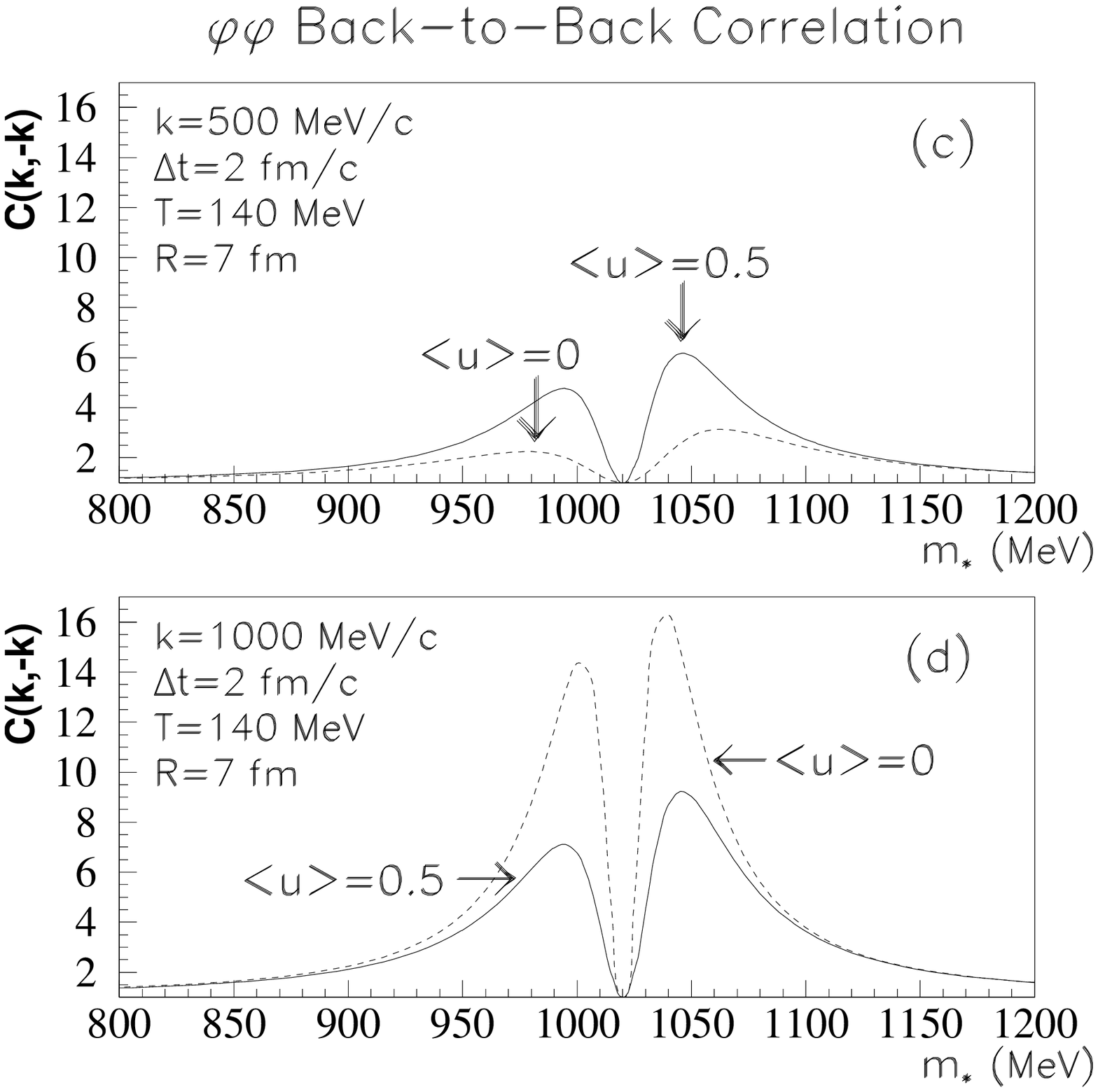}
 \caption{In the left panel, parts (a) and (b) illustrate the effect
of finite emission times. The dashed curves, corresponding to a
sudden emission ($\Delta t =0$), have been decreased by a factor
of $400$, and the solid curves show the suppression caused by a
finite emission duration, of about $\Delta t \simeq 2$~fm/c, which
drastically reduces the BBC magnitude. In the right panel we
illustrate the cases with and without flow, for two values of the
momentum of the back-to-back pair (in (c), $k=500$ MeV/c and in
(d), $k=1000$ MeV/c). The mass-shifting is supposed to occur in
the entire system volume. In the plots, $m_\star$ is the in-medium
modified mass and T stands for the freeze-out temperature. }
\end{figure}

For completeness, we write below the expression of the
Back-to-Back Correlation function for the case where the mass
shift occurs in entire volume of the system, $V$. A detailed
discussion and more complete formulation of the problem, including
the case of mass-shift in a smaller portion of the system, can be
found in Ref.\cite{phkpc05}. In what follows, we will concentrate
on the value of momenta of the participant pair that maximizes the
BBC signal, i.e., the case in which ${\mathbf k}_1 = - {\mathbf
k}_2 = {\mathbf k}$, using the fact that the single-inclusive
distribution depends only on the absolute value of the momentum,
i.e., $G_c({\mk},{\mk}) = G_c(-{\mk},-{\mk})$. Nevertheless, since
the strict condition of back-to-back pair holds only in the rest
frame of the medium, we implicitly are considering the case
corresponding to a weak flow coupling in the expanding system,
which is expected to be fulfilled for pair emitted near the center
of the system. In this case, the BBC correlation function can then
be written as
\bea && \!\!\!\!\!\!\!\!\!\!\!\! C^V_{BBC}({\mk},-{\mk}) = 1 +
\big| c_0 \; s_0 \big|^2 \,
\left[ 2 R^3 \left(1+\frac{m^2 <u>^2}{ m_\star T }\right)^{-3/2}
\exp \left( - \frac{m_\star}{T} - \frac{{\mathbf k}^2}{2 m_\star
T} \right) + R^3 \; \right]^2
\nonumber \\
&& \!\!\!\!\!\!\!\!\!\!\!\!\!\!\!\! \times \Bigg[\frac{ R^3 \left(
|c_0|^2 + |s_0|^2 \right)}{\left(1+\frac{m^2 <u>^2}{ m_\star T
}\right)^{3/2}} \exp \left( - \frac{m_\star}{T} - \frac{{\mathbf
k}^2}{2m_\star T} + \frac{m^2 <u>^2 {\mathbf k}^2}{2 m_\star^2
T^2(1+\frac{m^2 <u^2>}{ m_\star T })} \right) + R^3 |s_0|^2
 \Bigg]^{-2}\!\!\!. \label{BBCcorr1Vf}\eea

In Figure 2 we illustrate some of the results found in this
non-relativistic approximation, in the particular case of weak
flow coupling. We see that the cases corresponding to the absence
of flow and to its inclusion produce similar results within the
limits of our illustration. However, depending on the value of
$k_1=-k_2=k$, there are noticeable differences. Being so, we see
that, for smaller values of $k$, the presence of flow seems to
slightly enhance the signal, whereas at large values of $k$, the
non-flow case wins. Nevertheless, the non-flow case grows faster
with increasing $k$. In this particular example discussed here,
the effect is mainly due to the denominator in Eq.
(\ref{BBCcorr1Vf}), since it contains the flow parameter in the
exponential's coefficient.

\subsection{Summary and Conclusions}

Our main goal on presenting these new results here was to revive
the discussion on the search of the squeezed BBC. For fulfilling
this purpose, we estimated the strength of the BBC signal in a
more realistic situation, considering the mass-shifting in a
finite region, and the emission occurring during a finite time
interval. We also considered that the system expands
non-relativistically and analyzed the simplified situation of weak
flow dependence (i.e., back-to-back pairs emitted close to central
system region) of the BBC. Finally, we employed $\phi\phi$ pair
correlation for illustration. We showed in Figure 2 the
back-to-back correlation function versus the in-medium shifted
mass, $m_\star$. We also saw that both the non-flow and the flow
cases produced similar results, with pronounced maxima around $m
\approx m_\star$. Although we did not show here results
corresponding to the case of mass-shift occurring only in a
smaller portion of the system, it is shown
elsewhere\cite{phkpc05,QM2005} to decrease proportionally to the
size of the mass-shift region. However, the effect of decreasing
the system size is far less significant than the finite emission
time for reducing the BBC magnitude. We also saw that, at least in
its weak-coupling limit, the flow may work for slightly enhance
the BBC signal for small values of the momentum $k$. Our main
conclusion, however, is that, in the particular framework
discussed above, a sizeable strength of the squeezed BBC signal
could be seen, making it a promising effect to be searched for
experimentally at RHIC.

Naturally, for a more refined calculation, it is mandatory to
introduce a model-based mass-shift. After that, it is also
essential to perform more realistic calculations with flow, in a
less particular kinematic region, while simultaneously searching
for those which could optimize the observation of the BBC signal.
Also an estimate of the shape and width of the BBC around the
direction $k_1=-k_2=k$ should be implemented.
Finally, for being able to make predictions closer to the
experimental conditions, it will be extremely important to obtain
some feed-back on the experimental acceptance, conditions, and
restrictions that could finally lead to the BBC discovery.

\vspace{-.2cm}
\begin{theacknowledgments}
One of us (SSP) would like to express her gratitude to the Funda\c
c\~ao de Amparo \`a Pesquisa do Estado de S\~ao Paulo (FAPESP) for
the financial support (Proc. N. 05/52190-1 and 2004/10619-9 - {
\sl Projeto Tem\'atico}), which permitted her to participate in
the WPCF, and in the ISMD 2005. T. Cs\"org\H o was supported by
the grant OTKA T038406.
\end{theacknowledgments}

\end{document}


\endinput